\documentclass{aastex631}
\usepackage{longtable}
\usepackage{subfigure}
\usepackage{threeparttablex}

\begin{document}

\title{GRB 241030A: a prompt thermal X-ray emission component and diverse origin of the very early UVOT WHITE and U band emission}

\author[0000-0002-0786-7307]{Qiu-Li Wang}
\affiliation{Key Laboratory of Dark Matter and Space Astronomy, Purple Mountain Observatory, Chinese Academy of Sciences, Nanjing 210023, China}
\affiliation{School of Astronomy and Space Science, University of Science and Technology of China, Hefei 230026, China}

\author[0000-0003-2915-7434]{Hao Zhou}
\affiliation{Key Laboratory of Dark Matter and Space Astronomy, Purple Mountain Observatory, Chinese Academy of Sciences, Nanjing 210023, China}

\author[0000-0002-8385-7848]{Yun Wang}
\affiliation{Key Laboratory of Dark Matter and Space Astronomy, Purple Mountain Observatory, Chinese Academy of Sciences, Nanjing 210023, China}

\author[0000-0002-9037-8642]{Jia Ren}
\affiliation{Key Laboratory of Dark Matter and Space Astronomy, Purple Mountain Observatory, Chinese Academy of Sciences, Nanjing 210023, China}

\author[0000-0002-1481-4676]{Samaporn Tinyanont}
\affiliation{National Astronomical Research Institute of Thailand, 260 Moo 4, Donkaew, Maerim, Chiang Mai 50180, Thailand}

\author[0000-0003-3257-9435]{Dong Xu}
\affiliation{National Astronomical Observatories, Chinese Academy of Sciences, Beijing 100049, China}

\author{Ning-Chen Sun}
\affiliation{School of Astronomy and Space Science, University of Chinese Academy of Sciences, Beijing 100049, China}
\affiliation{National Astronomical Observatories, Chinese Academy of Sciences, Beijing 100049, China}
\affiliation{Institute for Frontiers in Astronomy and Astrophysics, Beijing Normal University, Beijing 102206,  China}

\author[0000-0002-8149-8298]{Johan P.U. Fynbo}
\affiliation{Cosmic Dawn Center (DAWN), Copenhagen 2200, Denmark.}
\affiliation{Niels Bohr Institute, Copenhagen University, Jagtvej 155, DK-2200, Copenhagen N, Denmark}

\author[0000-0002-7517-326X]{Daniele B. Malesani}
\affiliation{Cosmic Dawn Center (DAWN), Copenhagen 2200, Denmark.}
\affiliation{Niels Bohr Institute, Copenhagen University, Jagtvej 155, DK-2200, Copenhagen N, Denmark}

\author{Jie An}
\affiliation{National Astronomical Observatories, Chinese Academy of Sciences, Beijing 100049, China}
\affiliation{School of Astronomy and Space Science, University of Chinese Academy of Sciences, Beijing 100049, China}

\author{Rungrit Anutarawiramkul}
\affiliation{National Astronomical Research Institute of Thailand, 260 Moo 4, Donkaew, Maerim, Chiang Mai 50180, Thailand}

\author{Pathompong Butpa}
\affiliation{National Astronomical Research Institute of Thailand, 260 Moo 4, Donkaew, Maerim, Chiang Mai 50180, Thailand}

\author{Shao-Yu Fu}
\affiliation{National Astronomical Observatories, Chinese Academy of Sciences, Beijing 100049, China}

\author{Shuai-Qing Jiang}
\affiliation{National Astronomical Observatories, Chinese Academy of Sciences, Beijing 100049, China}
\affiliation{School of Astronomy and Space Science, University of Chinese Academy of Sciences, Beijing 100049, China}

\author{Xing Liu}
\affiliation{National Astronomical Observatories, Chinese Academy of Sciences, Beijing 100049, China}
\affiliation{School of Astronomy and Space Science, University of Chinese Academy of Sciences, Beijing 100049, China}

\author{Kritsada Palee}
\affiliation{National Astronomical Research Institute of Thailand, 260 Moo 4, Donkaew, Maerim, Chiang Mai 50180, Thailand}

\author{Pakawat Prasit}
\affiliation{National Astronomical Research Institute of Thailand, 260 Moo 4, Donkaew, Maerim, Chiang Mai 50180, Thailand}

\author[0000-0002-9022-1928]{Zi-Pei Zhu}
\affiliation{National Astronomical Observatories, Chinese Academy of Sciences, Beijing 100049, China}

\author[0000-0003-4977-9724]{Zhi-Ping Jin}
\affiliation{Key Laboratory of Dark Matter and Space Astronomy, Purple Mountain Observatory, Chinese Academy of Sciences, Nanjing 210023, China}
\affiliation{School of Astronomy and Space Science, University of Science and Technology of China, Hefei 230026, China}

\author[0000-0002-9758-5476]{Da-Ming Wei}
\affiliation{Key Laboratory of Dark Matter and Space Astronomy, Purple Mountain Observatory, Chinese Academy of Sciences, Nanjing 210023, China}
\affiliation{School of Astronomy and Space Science, University of Science and Technology of China, Hefei 230026, China}

\correspondingauthor{Hao Zhou, Yun Wang, Kaew S. Tinyanont, Zhi-Ping Jin}
\email{haozhou@pmo.ac.cn, wangyun@pmo.ac.cn, samaporn@narit.or.th, jin@pmo.ac.cn}

\begin{abstract}
We present a detailed analysis of the long-duration GRB 241030A detected by {\it Swift}. Thanks to the rapid response of XRT and UVOT, the strongest part of the prompt emission of GRB 241030A has been well measured simultaneously from optical to hard X-ray band. The time-resolved WHITE band emission shows strong variability, largely tracing the activity of the prompt gamma-ray emission, may be produced by internal shocks too. The joint analysis of the XRT and BAT data reveals the presence of a thermal component with a temperature of a few keV, which can be interpreted as the photosphere radiation, and the upper limit of the Lorentz factor of this region is found to range between approximately 20 and 80.
The time-resolved analysis of the initial U-band exposure data yields a very rapid rise ($ \sim t^{5.3}$) with a bright peak reaching 13.6 AB magnitude around 410 seconds, which is most likely attributed to the onset of the external shock emission. The richness and fineness of early observational data have made this burst a unique sample for studying the various radiation mechanisms of gamma-ray bursts.

% This example manuscript is intended to serve as a tutorial and template for
% authors to use when writing their own AAS Journal articles. The manuscript
% includes a history of \aastex\ and includes figure and table examples to illustrate these features. Information on features not explicitly mentioned in the article can be viewed in the manuscript comments or more extensive online
% documentation. Authors are welcome replace the text, tables, figures, and
% bibliography with their own and submit the resulting manuscript to the AAS
% Journals peer review system.  The first lesson in the tutorial is to remind
% authors that the AAS Journals, the Astrophysical Journal (ApJ), the
% Astrophysical Journal Letters (ApJL), the Astronomical Journal (AJ), and
% the Planetary Science Journal (PSJ) all have a 250 word limit for the 
% abstract\footnote{Abstracts for Research Notes of the American Astronomical 
% Society (RNAAS) are limited to 150 words}.  If you exceed this length the
% Editorial office will ask you to shorten it. This abstract has 161 words.

\end{abstract}


%% and \citett commands to identify citations.  The citations are

\section{Introduction} \label{sec:intro}

%% Also note that the akcnowlodgment environment does not support long amounts of text. If you have a lot of people and institutions to acknowledge, do not use this command. Instead, create a new \section{Acknowledgments}.

Gamma-ray bursts (GRBs) are the most powerful explosions in the universe. Several models like dissipative photosphere, internal shock and magnetic dissipation have been developed to explain the nature of prompt emission, and the afterglow emission is thought to originate from the external shock composed of a forward shock (FS) and a reverse shock (RS), which were only detected in some GRBs \citep{Mészáros_1997,Mészáros_1999,Sari_1999}. 

Only a few optical observations have been made contemporaneous with gamma-ray bursts, due to the limitation of the response speed of optical instruments in the past. The first optical flash discovered in the prompt phase is GRB 990123 \citep{akerlof_observation_1999}. Its bright emission in the optical band can be explained by the reverse shock of a thin shell \citep{2002ChJAA...2..449F,2003ApJ...595..950Z}. The rapid response and precise localization capabilities of the \emph{Swift} mission \citep{Gehrels_2004} make it possible to obtain simultaneous data from the optical band to the $\gamma$-ray band. Furthermore, Space-based multi-band astronomical Variable Objects Monitor (SVOM), with four onboard instruments launched in 2024, has a larger field-of-view (FOV) and higher sensitivity which enhances the possibility to observe GRB from the optical band to the $\gamma$-ray band at an early stage \citep{2014SPIE.9144E..24G, Wei2024, 2024RAA....24k5006W}. Several models have been developed to explain bright optical flashes of GRBs at the early stage. For example, the early-time optical emission could be the onset of the afterglow emission (either from FS or RS), and some argue early optical flashes of GRBs can usually arise from the internal shock \citep{2007MNRAS.374..525W}. Some physical properties of GRB ejecta can be constrained from very early afterglow emission data, e.g., the magnetized degree of the outflow powering GRBs \citep{2004A&A...424..477F,2005ApJ...628..315Z}. Hence, a systematic study of multi-band emission of GRBs at the early stage is helpful to constrain GRB emission models.

In this article, we report the analysis of long GRB 241030A data obtained by the Neil Gehrels \emph{Swift}. The {\it Swift} burst alert telescope \citep[BAT, ][]{barthelmy_burst_2005} was triggered by a precursor of GRB 241030A at 100 seconds before the main burst, leading to {\it Swift} performed multi-band observations of the prompt phase with high temporal resolution, using the X-ray telescope \citep[XRT, ][]{burrows_swift_xrt} and the ultraviolet/optical telescope \citep[UVOT, ][]{roming_swift_2005}. Similar to \cite{2023NatAs...7.1108J} and \cite{2023ApJS..268...65Z}, we split the UVOT data taken in the event mode to short time bins to explore the optical temporal behavior at the early stage of the burst. The prompt $\gamma$-ray emission of this burst exhibits several pulses overlapped with each other. Similar temporal behavior is also seen in the optical and X-ray data. After the prompt phase, the U-band data shows a bump peaking around 410\,s after the trigger, which is attributed as the onset of the afterglow. The early behavior of the afterglow is well sampled, which sets good  constraints for parameters of the standard fireball model.

We carried out ground-based optical follow-up using the Thai Robotic Telescope (TRT) and the Nordic Optical Telescope (NOT), observations and the reduction of the data are described in Sec \ref{sec:observations}. The prompt emission analysis and board-band spectral fit results are presented in Sec \ref{sec:prompt}, and the afterglow analysis are presented in Sec \ref{sec:afterglow}. In Sec \ref{sec:summary}, we summarize and discuss our work. The standard cosmology model with $H_{0}=67.4$ km s$^{-1}$Mpc$^{-1}$, $\Omega_{M}=0.315$ and $\Omega_{\Lambda}=0.685$ \citep{2020A&A...641A...6P} is adopted in this paper. All errors are given at the 1$\sigma$ confidence level unless otherwise stated.

\section{OBSERVATIONS AND DATA REDUCTIONS} \label{sec:observations}

\subsection{Observations} \label{subsec:subsection1}
At 05:48:03 UT on October 30, 2024 ($T_{0}$), GRB 241030a simultaneously triggered Fermi-GBM and {\it Swift}/BAT \citep{2024GCN.37955....1F,2024GCN.37956....1K}. Based on the BAT localization, {\it Swift} immediately slewed to the source and initiated follow-up observations. Then a bright, uncatalogued X-ray source was identified by XRT 73.5 s after BAT trigger at the location RA = 22h 52m 33.30s, DEC = +80$\,^\circ$ 26$\,\arcmin$ 59.6$\,\arcsec$ (J2000) with a 90\% uncertainty of 2.2$\,\arcsec$ \citep{2024GCN.37962....1B}. UVOT began to observe the field of GRB 241030A about 83 seconds after the BAT trigger in the event mode \citep{2024GCN.37974....1B}. In addition, Fermi-LAT also detected GeV photons from this event \citep{2024GCN.37979....12}. Furthermore, SVOM/VT observed the field of GRB 241030A via ToO observations started at 07:03:11 UTC, about 1.25 hours after the burst \citep{2024GCN.37965....1S}. The VT conducted observations simultaneously in two channels: VT\_B (400nm-650nm) and VT\_R (650nm-1000nm). SVOM/C-GFT start to observe the field of GRB 241030A on 09:54:53 UT, about 4.09 hours after the trigger in the commissioning phase \citep{2024GCN.37970....1S}. A series of g, r and i band images were obtained with exposure time of 30 seconds. The SVOM/GRM was triggered in-flight by GRB 241030A at 05:48:14 U \citep{2024GCN.37972....1S}. However, at the time of the burst ECLAIRs was not collecting data. In addition, the optical counterpart of GRB 241030A was observed with the Low
Resolution Imaging Spectrometer on the Keck I 10 m telescope, revealing a continuum spectrum and narrow absorption lines, with a likely redshift of 1.411 \citep{2024GCN.37959....1Z}.

%% initial draft added by AJ
Approximately 26 minutes after $T_{0}$, we manually triggered the 70 cm telescopes at Sierra Remote Observatories in the USA, one of the nodes of the worldwide Thai Robotic Telescope (TRT), obtaining continuous and complete light curve with 60 s, 90 s, and 180 s frames in the R band that lasted until three hours after the trigger time. About three days later, we acquired further observations in the Sloan g, r, and z band using the Nordic Optical Telescope (NOT; 2.56 m at the Roque de los Muchachos observatory, La Palma, Spain) telescope.
%% initial draft added by AJ

\subsection{Data reduction} \label{subsec:subsection2} 
\subsubsection{Swift Data} \label{subsubsec:subsubsection1}

The data from {\it Swift} BAT and UVOT, was reduced with the HEASoft 6.34. The BAT light curve (15 - 350\,keV) from $T_0-60$\,s to $T_0+242$\,s was extracted with a time bin size of 1\,s. The XRT data was reduced with the online {\it Swift}/XRT data products generator\footnote{\url{https://www.swift.ac.uk/burst_analyser/}}\citep{ev2007, ev2009}. UVOT observed GRB 2410301A in the V, B, U, W1, M2, W2 and WHITE bands for several epochs. For data taken in the image mode, we started from the level 2 UVOT products and made photometry with the standard aperture. For the WHITE-band and U-band data taken in the event mode, a fixed time bin of 5\,s (10\,s) was applied to extract the light curve in the WHITE (U) band. The standard aperture is also adopted to make photometries of the event data. The UVOT photometries are listed in Table \ref{table:1} .

\begin{figure}[!ht]
    \centering
    \includegraphics[width=0.5\textwidth]{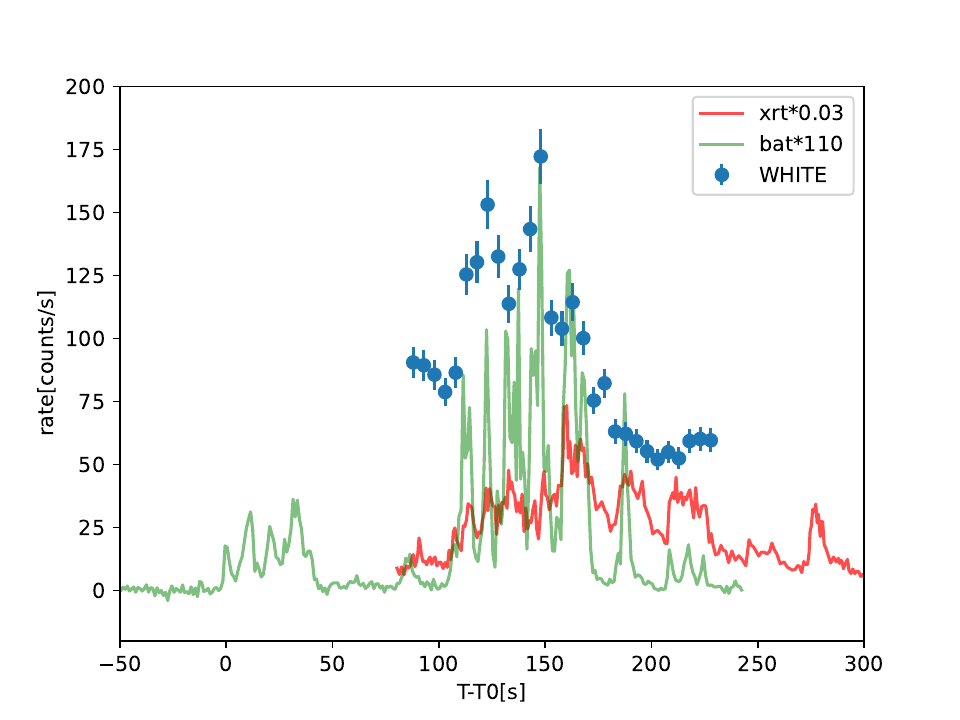}
    \caption{Mulit-band prompt emission light curves of GRB 241030A. Data are collected by {\it Swift} BAT (green), XRT (red), and UVOT (blue). Note that the UVOT data are all in WHITE band.}
    \label{fig:1}
\end{figure}

As shown in the Figure \ref{fig:1}, the BAT light curve showed a complex structure with many overlapped pulses that lasted for about 250\,s. Generally, precursors are broadly defined as fainter emissions followed by a quiescent period before the main emission of the GRB \citep{2022AdSpR..70.1512B}. There is a quiescent period from $\sim50$ to $\sim80$\,s after the BAT trigger, which suggests the first group of the $\gamma$-ray pulses could be the precursor emission. Compared with the second group of $\gamma$-ray pulses starting from $\sim T_0+100$\,s as well as the precursor, the intensity of the small pulse from $\sim T_0+80$ to $T_0+90$\,s is weak and followed by a short quiescent period lasting for about 10\,s implying the small pulse also could be the precursor emission. Hence, we suggest that GRB 241030A is a burst triggered by its precursor emission, and the true start time of the main prompt emission is about $T_0+100$\,s, which is similar to another {\it Swift}/BAT GRB 060124 in the literature \citep{2006A&A...456..917R}. The X-ray light curve also shows about ten pulses which almost trace the $\gamma$-ray light curve. The light curve in the WHITE band shows 5 peaks and the brightest one is around 148\,s. Only 4 optical peaks coincide with $\gamma$-ray peaks. The optical peak around 180\,s does not coincide with any $\gamma$-ray peak but with a small X-ray peak. The similarity between the UV/optical, X-ray and $\gamma$-ray light curves suggests that in the prompt phase, the UV/optical, X-ray and $\gamma$-ray emission may have the same origin (e.g., activities of the central engine or the internal shock). The U-band light curve is smoother than the WHITE-band light curve and clearly shows a bump at $\sim410$\,s with a steep rise and a normal decay which may originate from external shock region.

\subsubsection{Fermi Data} \label{subsubsec:subsubsection2}
Fermi-GBM is composed of 12 NaI(TI) detectors and 2 BGO detectors \citep{2009ApJ...702..791M}.
Based on the angle between the source location and the pointing direction of each detector, and considering the poor quality of the BGO observation data, we selected only the n0 detector data for subsequent joint spectral analysis. And we used {\tt GBM data tools} \citep{GbmDataTools} to generate the files required for spectral analysis from the TTE (Time-Tagged Event) data. 

In order to produce the GeV light curves, we also performed data reduction for the Fermi-LAT data. Fermi-LAT is a high-performance gamma-ray telescope designed for a photon energy range from 30 MeV to 1 TeV \citep{2009ApJ...697.1071A}. For the Fermi-LAT data reduction, we selected photon events within the energy range of 100 MeV to 100 GeV, using the {\tt TRANSIENT} event class and the {\tt FRONT+BACK} type. To minimize contamination from the Earth’s limb, photons with zenith angles exceeding 100$^{\circ}$ were excluded. Subsequently, we identified good time intervals by applying the quality filter condition ({\tt (DATA\_QUAL$>$0 \&\& LAT\_CONFIG==1)}). In the standard unbinned likelihood analysis procedure\footnote{\url{ https://fermi.gsfc.nasa.gov/ssc/data/analysis/scitools/}}, we selected a 10$^{\circ}$ region centered on the location of GRB 241030A as the region of interest (ROI). The initial model for the ROI region, generated by the {\tt make4FGLxml.py} script\footnote{\url{https://fermi.gsfc.nasa.gov/ssc/data/analysis/user/}}, includes the galactic diffuse emission template ({\tt gll\_iem\_v07.fits}), the isotropic diffuse spectral model for the {\tt TRANSIENT} data ({\tt iso\_P8R3\_TRANSIENT020\_V3\_v1.txt}) and all the Fourth Fermi-LAT source catalog~\citep[{\tt gll\_psc\_v35.fit};][]{2023arXiv230712546B} sources. We defined GRB 241030A as a point source in the model file, with the model set to {\tt PowerLaw2}\footnote{\url{https://fermi.gsfc.nasa.gov/ssc/data/analysis/scitools/source_models.html}}. And an automatic rebinning algorithm was employed, ensuring at least 10 photons per bin and a TS $>$ 9 to guarantee that the flux in each bin is significant.
%For the Fermi-GBM observation, we selected the GBM-TTE data with a temporal resolution of 1.024\,s of the brightest detector: NaI\#0 according to the SNR. All data reductions like time slicing , background spectrum extraction were using standard procedures with Fermi GBM Data Tools (version 1.1.1, \cite{GbmDataTools}). Channels with energy from 8 to 900\,keV were adopted for NaI. 

%\subsubsection{Joint Spectral Fitting} \label{subsubsec:subsubsection3}
%The joint spectral fitting, including XRT (0.3 - 10 keV), BAT (15 - 150 keV), GBM/NaI (8 - 900 keV) and optical data, was performed with {\tt XSPEC} (version 12.14.1). We sliced the prompt emission epoch ($T_{0}+85$ s to $T_{0}+230$ s) into 29 time intervals with a time bin size of 5 s aligned with the optical observation intervals. Given the suboptimal SNR of less than 3 in the GBM data at certain epochs, the GBM data within specific time intervals (105 to 175 seconds, 185 to 195 seconds, and 215 to 220 seconds) was only adopted for spectral fitting analysis. In the afterglow phase, spectral energy distributions (SEDs) were constructed at $T_{0}+755$ s when there is the best coverage of multi-band data from optical to X-ray band. We jointly fit board-band data with the $\chi^{2}$ statistic, except GBM data, for which the {\tt pgstat} (i.e., Poisson data with Gaussian background) statistic was used.

\subsubsection{Ground-base Photometry} \label{subsubsec:subsubsection3}
After conducting standard data reduction with the Image Reduction and Analysis Facility \citep{tody_iraf_1986} and performing astrometric calibration using Astrometry.net \citep{lang_astrometrynet_2010}, the apparent photometry was calibrated against the Pan-STARRS1 DR2 catalog \citep{flewelling_pan-starrs1_2020}. The Johnson-Cousins filters were calibrated using magnitudes that were transformed from the Sloan system. \footnote{\url{https://live-sdss4org-dr12.pantheonsite.io/algorithms/sdssUBVRITransform/\#Lupton>}} The photometric results without Galactic extinction correction are tabulated in Table \ref{table:ground obs}.
%% very old-school reduce process, or no need to write?

\section{PROMPT EMISSION ANALYSIS} \label{sec:prompt}
\subsection{Broadband light curves of the prompt emission} \label{subsec:subsection31}

XRT and UVOT began to observe GRB 241030A at $T_{0}+73.5$\,s and $T_{0}+83$\,s, respectively. Both XRT and UVOT obtained well-sampled data during the prompt phase for about 150\,s. Combining the $\gamma$-ray data observed by BAT (15.0 keV to 150 keV) and GBM data (8 keV to 900 keV), the entire prompt phase ($\sim T_0+75$\,s to $\sim T_0+250$\,s) is almost covered by multi-band observations, which makes it possible to study the temporal and spectral properties of the prompt emission in detail.

Figure \ref{fig:1} shows multi-band light curves in the prompt phase. BAT data shows two groups of several peaks from $T_{0}$ to $T_{0}+50$\,s and $T_{0}+80$\,s to $T_{0}+230$\,s with a peak at $T_{0}+148$\,s. We consider the first group of peaks in BAT data as the precursor emission, which triggered BAT, and the second group is the prompt emission. XRT data show overlapped peaks and the brightest one is at $T_{0}+160$ s. The UVOT WHITE-band light curve in the prompt phase is binned with a temporal size of 5\,s, and shows three major peaks, which seems to be correlated with the BAT data. Optical data with high temporal resolution provides clear evolution of the optical emission in the prompt emission. The brightest optical peak occurs at $\sim T_{0}+148$\,s when the brightest $\gamma$-ray peak occurs. The other 2 optical peaks occur at $\sim T+123$\,s and $\sim T+163$\,s when the $\gamma$-ray light curve also peaks. Generally speaking, optical data exhibit complicated variability and partially trace the gamma-ray data, implying that they share the same origin and are mainly produced by internal shocks.

\begin{figure}[!hb] 
\centering  
\subfigbottomskip=1pt
\subfigure[PL with BB]
{
	\label{sub.1}
	\includegraphics[width=0.35\linewidth]{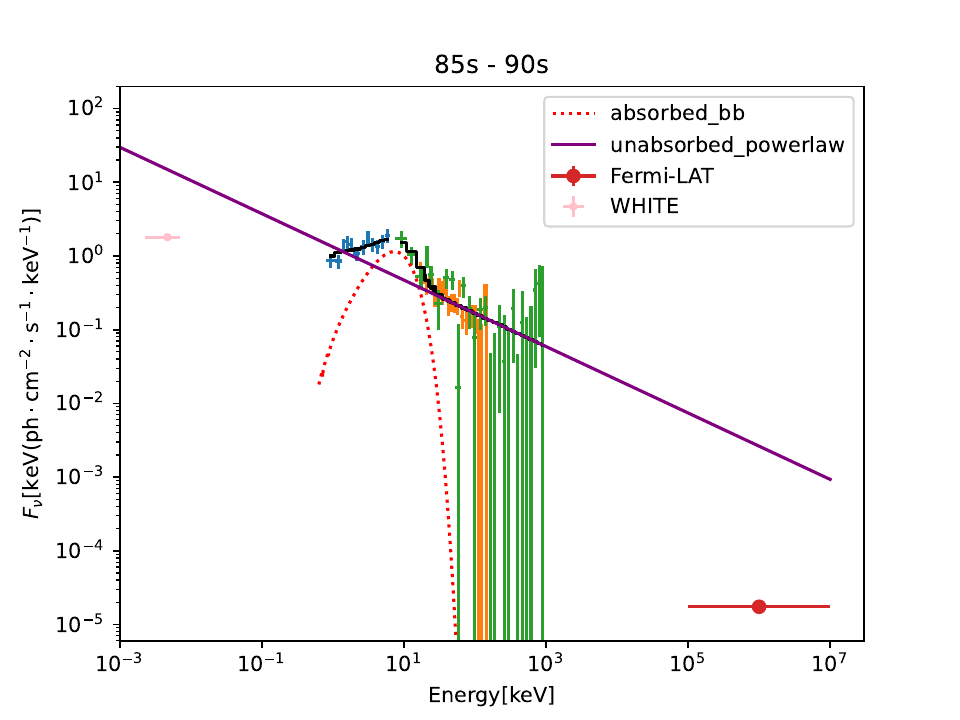}
}
\subfigure[PL only]
{
		\label{sub.2}
		\includegraphics[width=0.35\linewidth]{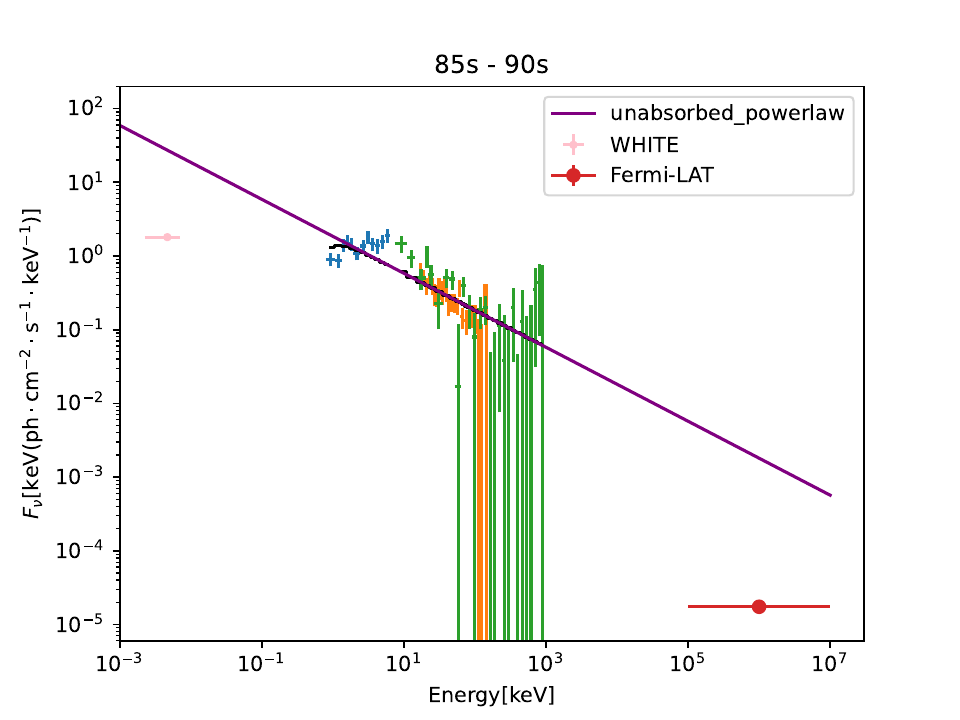}
	}

\subfigure[CPL with BB]
	{
		\label{sub.3}
		\includegraphics[width=0.35\linewidth]{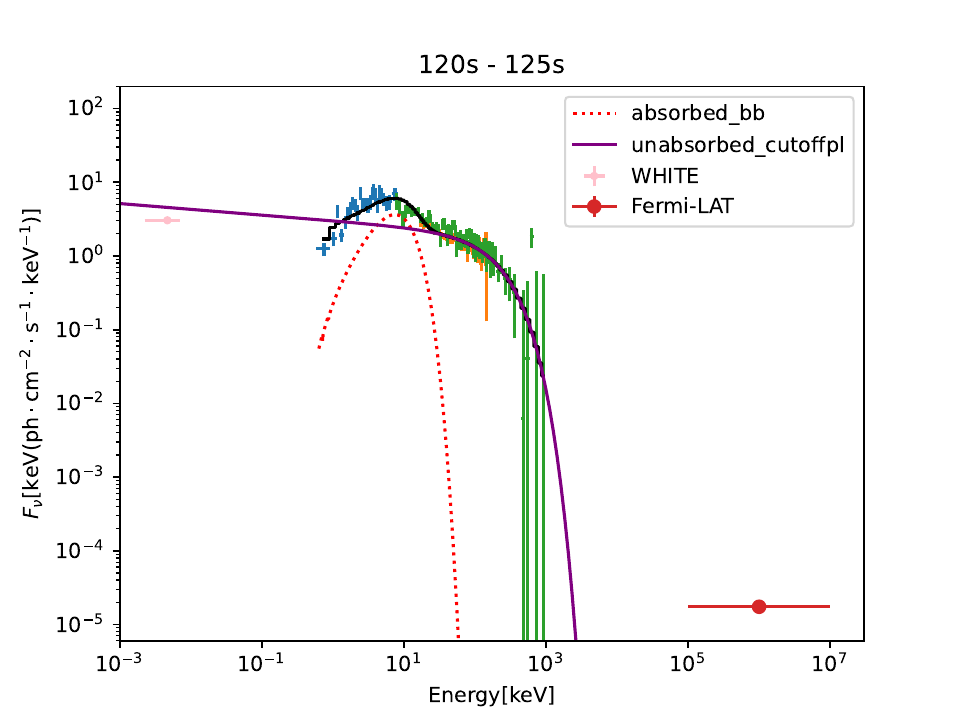}
	}
\subfigure[CPL only]
	{
		\label{sub.4}
		\includegraphics[width=0.35\linewidth]{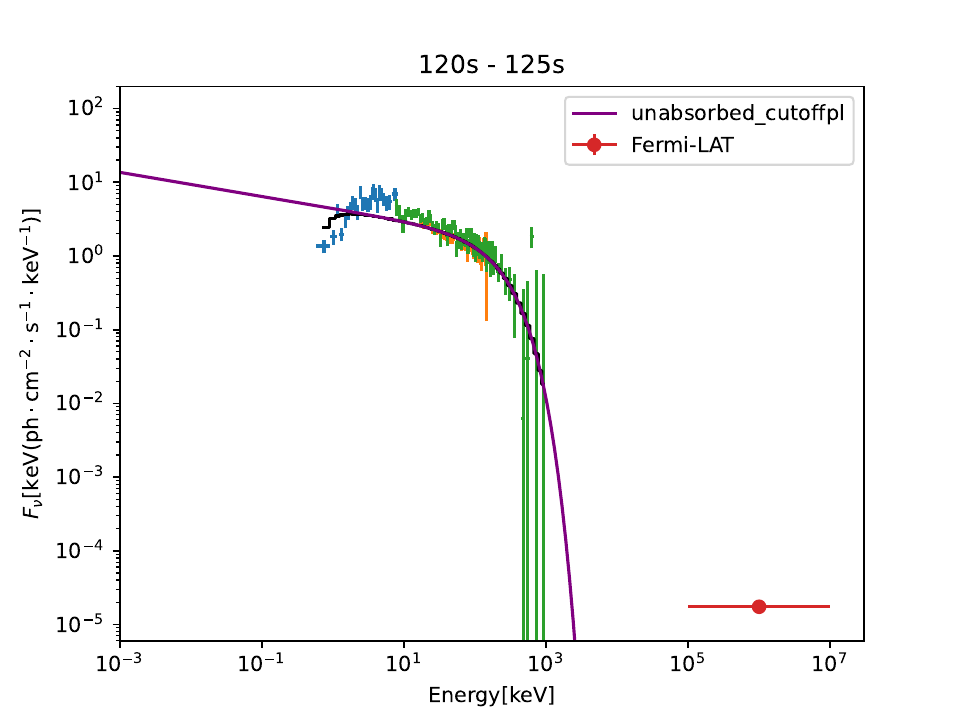}
	}
	\caption{The SEDs of GRB 241030A prompt emission. (a) PL with BB: the spectral fitting of 85 s to 90 s interval in PL with BB model. (b) PL only: the spectral fitting of 85 s to 90 s interval in PL only model. (c) CPL with BB: the spectral fitting of 120 s to 125 s interval in CPL with BB model. (d) CPL only: the spectral fitting of 120 s to 125 s interval in CPL only model. Pink points represent observed optical flux density in WHITE band and red points represent Fermi-LAT data. Due to the low SNR of the Fermi-LAT data, here we used the average flux density over the 0-260 seconds within the energy range of 0.1-10 GeV. Dotted red lines represent absorbed BB component and purple solid lines represent unabsorbed PL or CPL component.}
	\label{fig:bb}
\end{figure}

\subsection{Spectral analysis of the prompt emission} \label{subsec:subsection32}
The joint spectral fitting, including XRT (0.3 - 10 keV), BAT (15 - 150 keV), GBM/NaI (8 - 900 keV) and optical data, was performed with {\tt XSPEC} (version 12.14.1). We sliced the prompt emission epoch ($T_{0}+85$ s to $T_{0}+230$ s) into 29 time intervals with a time bin size of 5 s aligned with the optical observation intervals. Given the suboptimal signal-to-noise ratio (SNR) of less than 3 in the GBM data at certain epochs, the GBM data within specific time intervals (105 to 175 seconds, 185 to 195 seconds, and 215 to 220 seconds) was only adopted for spectral fitting analysis. We jointly fit board-band data with the $\chi^{2}$ statistic, except GBM data, for which the {\tt pgstat} (i.e., Poisson data with Gaussian background) statistic was used. Similar to \citep{Wang_2023}, four models with Galactic absorption ($tbabs$) and host galaxy ($ztbabs$) absorption were used here for spectral fitting, including power-law (PL), cutoff power-law (CPL), power-law with blackbody (PL+BB) and cutoff power-law with blackbody (CPL+BB). The Galactic equivalent hydrogen column density $N_{H}$ is fixed as $1.79\times 10^{21}$\,cm$^{-2}$ \citep{2024GCN.37988....1A} and the host galaxy equivalent hydrogen column density could be ignored according to the fitting result of late-time ($>T_0+4500$\,s) averaged XRT spectrum. For some spectral energy distributions (SEDs), the SNR for $E\gtrsim100$\,keV is low and the cut energy of the CPL model is poorly constrained. Hence, the PL model is adopted to fit the data instead of CPL.

As shown in Figure \ref{fig:bb}, the X-ray to $\gamma$-ray SEDs clearly show excesses around $\sim10$\,keV when fitted with a PL or CPL model, and the excesses can be fitted by a thermal component. Compared to models that employ only CPL or PL, our analysis reveals that models incorporating both CPL+BB or PL+BB have better fitting results. Models with BB exhibit lower values of the Bayesian Information Criterion (BIC) compared to those without. BIC is defined as follow: 
\begin{equation}
    BIC = -2\ln_{}{L} +k\ln_{}{N} 
\end{equation}
where $L$ is the maximized value of the likelihood function of the model, $k$ is the number of free parameters, $N$ is the number of data points. Table \ref{table:prompt_fit} shows detail parameters of spectral fit and differences of BIC, $\Delta_{\rm BIC}$, between models with BB and models without BB in different time intervals. Except for the epoch 140 - 145 s where $\Delta_{\rm BIC}$ = 2.97, $\Delta_{\rm BIC} > 10$ in all other epochs supports the existence of an additional thermal component and the $\Delta_{\rm BIC} $ exceeds 30 in 20 of 29 epochs which is strong evidence of an additional thermal component. For the case in 140 - 145 s, the flux of blackbody derived from the fitted blackbody parameters is very low, so it may be covered by CPL or PL component, which is also reasonable. 

\subsection{Parameters of the Photospheric Radiation}
\begin{figure}[tbhp]
    \centering
    \includegraphics[width=0.6\textwidth]{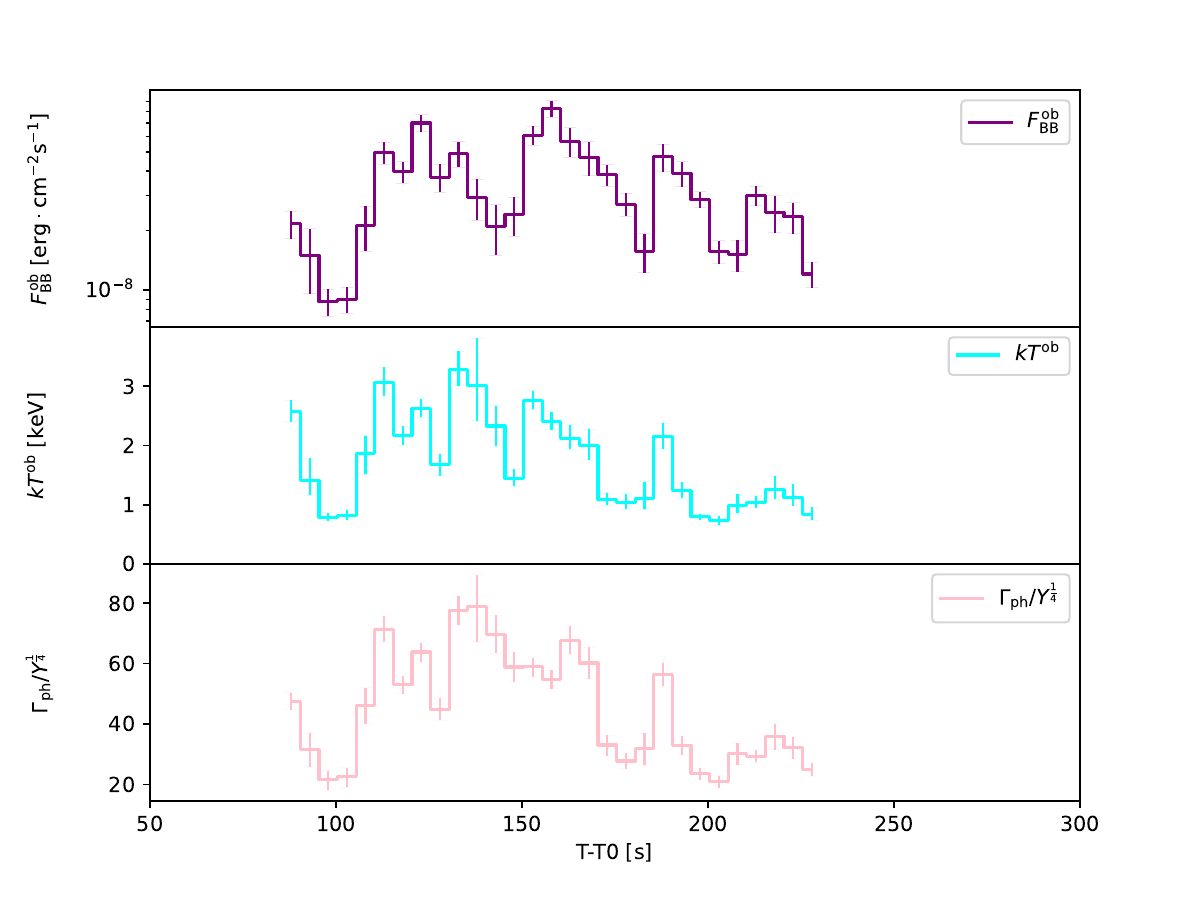}
    \caption{The evolution of different parameters for the thermal component in GRB 241030A. From top to bottom are the observed flux of the thermal component $F_{\rm BB}^{\rm ob}$, the observed temperature of the thermal component $kT^{\rm ob}$ and the Lorentz factor of the photosphere $\Gamma_{\rm ph}$ in the prompt emission, respectively. }
    \label{fig:lorentz}
\end{figure}

With the best fitted temperature $kT_{BB}$ and the flux $F_{BB}$ of the thermal component, we can estimate physical parameters of the fireball model, e.g., the Lorentz factor of photosphere using the method in \cite{Pe’er_2007}. The Lorentz factor of the photosphere $\Gamma_{\rm ph}$ is:
\begin{equation}\label{equ:Gamma}
\Gamma_{\rm ph}=\left[(1.06)(1+z)^{2}d_{L}\frac{YF^{ob}\sigma _{T}}{2m_{p}c^{3}\mathcal{R} } \right]^{\frac{1}{4} }
\end{equation}
where $d_{L}$ is the luminosity distance, $F^{ob}$ is the total observed flux of thermal and non-thermal components, $m_{p}$ is the mass of the proton, $\sigma_{T}$ is the Thomson scattering cross-section. In Pe'er's method, $Y = \epsilon / \epsilon_{\gamma}\ge 1$ is the ratio between the total fireball energy and the energy emitted in $\gamma$-ray. In our work, $Y = \epsilon / \epsilon_{\rm ob}$ is the ratio between the total fireball energy and the observed energy emitted in 0.3 - 150 keV here we exhibit the upper limit of $\Gamma_{\rm ph}$. $\mathcal{R}$ is the ratio factor defined as:
\begin{equation}\label{equ:R}
    \mathcal{R} \equiv \left( \frac{F^{ob}_{BB}}{\sigma T^{ob^{4}}}\right)^{\frac{1}{2} }
\end{equation}
where $\sigma$ is Stefan-Boltzmann constant, $T^{ob}$ and $F^{ob}_{BB}$ are the observed temperature and flux of the thermal component. Combing Equation (\ref{equ:Gamma}) and Equation (\ref{equ:R}), the Lorentz factor of photosphere $\Gamma_{\rm ph}$ can be calculated and the evolution of $\Gamma_{\rm ph}/Y^{\frac{1}4{}}$ is shown in Figure \ref{fig:lorentz}.

\section{AFTERGLOW EMISSION ANALYSIS} \label{sec:afterglow}
\subsection{Temporal behavior of the afterglow}
UV/optical light curves of the afterglow were phenomenologically fitted with a broken power-law (BPL) function, which has been widely used to fit afterglow light curves for both the rising and decay phases (e.g., \citealp{2007ApJ...670..565L,2012ApJ...758...27L,2015ApJS..219....9W,2018ApJ...859..163H}). The best fitted BPL model of the afterglow light curve could provide an initial understanding of the physical properties of this burst. The BPL can be written as:
\begin{equation}
f(t) = A \left[\left(\frac{t}{t_{b}}\right)^{\alpha_{{1}}\Delta}+\left(\frac{t}{t_{b}}\right)^{\alpha_{2}\Delta}\right]^{-1/\Delta}
\end{equation}
where $A$ is the normalization factor, $t_{b}$ is the break time, $\alpha_1$ and $\alpha_2$ are the temporal decay indices, $\Delta$ is the smooth parameter. 

Since the $\gamma$-ray emission from $T_0$ to $T_0+100$\,s could be the precursor emission, so the $T_0+100$\,s is treated as the true start time of the burst (i.e., the launch time of the jet). The best fitted BPL model of UV/optical light curves gives $\alpha_{1} = -5.3^{+0.4}_{-0.3}$ and $\alpha_{2} = 1.3^{+0.1}_{-0.1}$. Note that $\alpha_2$ is well constrained by 7 UVOT UV/optical bands. Hence, to account for the behavior of afterglow in the brightening phase, both RS and FS components should be considered, which is discussed in Section \ref{subsec:subsection42}.

\subsection{The spectral analysis of the afterglow } \label{subsec:subsection41}
The epoch $T_0+755$\,s is selected to construct the broad band SED since we have the best multi-band coverage data. To ensure a sufficient SNR, the XRT data from 706 to 806\,s was collected to create the X-ray spectrum, with the equivalent photon arrival time of $\sim 755$\,s. As shown above, all 7 UVOT UV/optical light curves in the decay phase can be well described by a power-law decay with a temporal decay index $\alpha_2=1.3^{+0.1}_{-0.1}$. Hence, the UV/optical data was interpolated to $T_0+755$\,s. The interpolated UV/optical fluxes used to construct the broad band SED are listed in Table \ref{table:2}. 

\begin{figure}[!ht]
    \centering
    \includegraphics[width=0.5\textwidth]{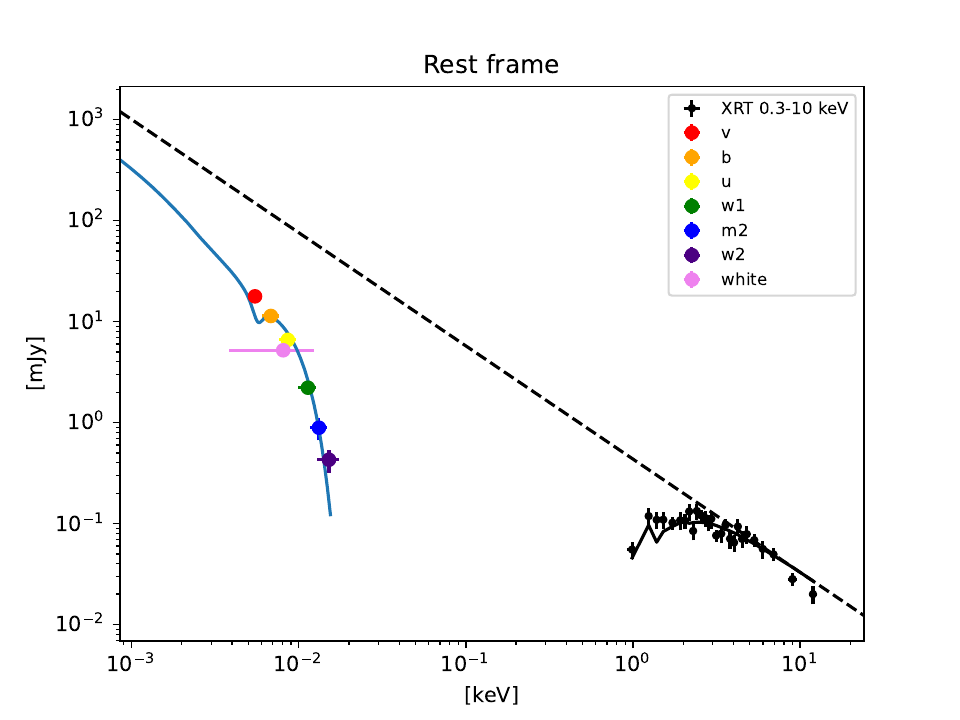}
    \caption{The broad band SED of GRB 241030A afterglow at 755\,s after the BAT trigger. Black points are observed data of XRT from 0.3 keV to 10.0 keV. Colorful points are the observed data in different UV/optical bands. Blue and black solid line represent the predicted absorbed model. Dashed line is the unabsorbed power-law model with a spectral index of $1.12$.}
    \label{fig:sed}
\end{figure}

The broad band SED was fitted with an absorbed PL model $F_\nu\propto\nu^{-\beta}$. The Milky Way extinction model with $E(B-V)= 0.1148$ \citep{1999PASP..111...63F} and $R_{V}=3.08$ is adopted. The $N_H$ from photoionization absorption model is set to $1.79\times 10^{21}\,{\rm cm}^{-2}$ \citep{2000ApJ...542..914W}. For the host, the extinction model of Small Magellanic Cloud model is adopted. The best fitted result gives the spectral index $\beta=1.12^{+0.02}_{-0.01}$ as predicted by the FS,  the the $E(B-V)$ of the host galaxy is $0.19^{+0.01}_{-0.01}$, which is shown in Figure \ref{fig:sed}.
During the Xspec fitting, the $tbabs$ and $ztbabs$ models are set as same as mentioned above. We tried both simple power-law and cut-off power-law model, however, we cannot use the cut-off power-law model here because the cutoff is in the $\gamma$-ray band and we don't have that detection at this epoch.

\subsection{Afterglow modeling} \label{subsec:subsection42}
After the prompt emission, the optical light curve is characterized by a steep rising to a peak around 410\,s. The beginning of steep rise is not observed by the UVOT, but it can be constrained that the start time of the rising is after 228 s according to the last data in WHITE band. The peak reaches U = 13.6 AB mag, and following the peak, UV/optical light curves decay monotonically without significant color evolution.

Meanwhile, the X-ray data show a different temporal behavior. After the prompt emission, the X-ray light curve has several pulses from 230 s to 300 s and peaks around 278 s. After the peak, the X-ray light curve shows a monotonic decay with a temporal decay index of -1.2. After $10^{5}$ s, a further steepening break in X-ray band is observed which is probably caused by a jet break.

The interaction between the relativistic outflow of the GRB and the circumburst medium induces the formation of a pair of relativistic shocks, i.e., FS and RS components. Initially the light curve is dominated by the RS emission characterized by a steep rise and a rapid decay. The FS takes over soon after the U-band peak, and dominates the decay phase. The electrons accelerated by these shocks are capable of generating multi-band emissions via synchrotron radiation and inverse Compton scattering. Numerous computational paradigms for modeling afterglows have been developed, we use the Python-wrapped Fortran package \citep{2024ApJ...962..115R} to perform numerical calculations of the radiation contributed from the FS component under the various effects. And the calculation of the contribution from the RS in the early afterglow is referenced from \citep{2013ApJ...776..120Y}. Therefore, under this model, the flux density at a certain time and frequency is:
\begin{equation}
    F_{t, \nu}=F(t,\nu, \Gamma_0,\epsilon_{e,f},\epsilon_{B,f},\theta_j,E_{\rm k,iso},p_{f},\epsilon_{e,r},\epsilon_{B,r},p_{f},n_0),
    \label{eq:fs_rs}
\end{equation}
where $\Gamma_0$ is the initial Lorentz factor, $\epsilon_e$ is the fraction of the shock energy into the energy of relativistic electrons, $\epsilon_b$ is the fraction of the shock energy converted into the energy of magnetic field, $\theta_j$ is the half-opening angle of the jet in radians, $E_{\rm k,iso}$ is the isotropic kinetic energy of the jet, $p$ is the electron energy distribution index, $n_0$ is the number density of the circumburst medium. And the additional subscripts $f$ and $r$ represent the FS and RS, respectively.

\begin{figure}[!ht]
    \centering
    \includegraphics[width=0.8\textwidth]{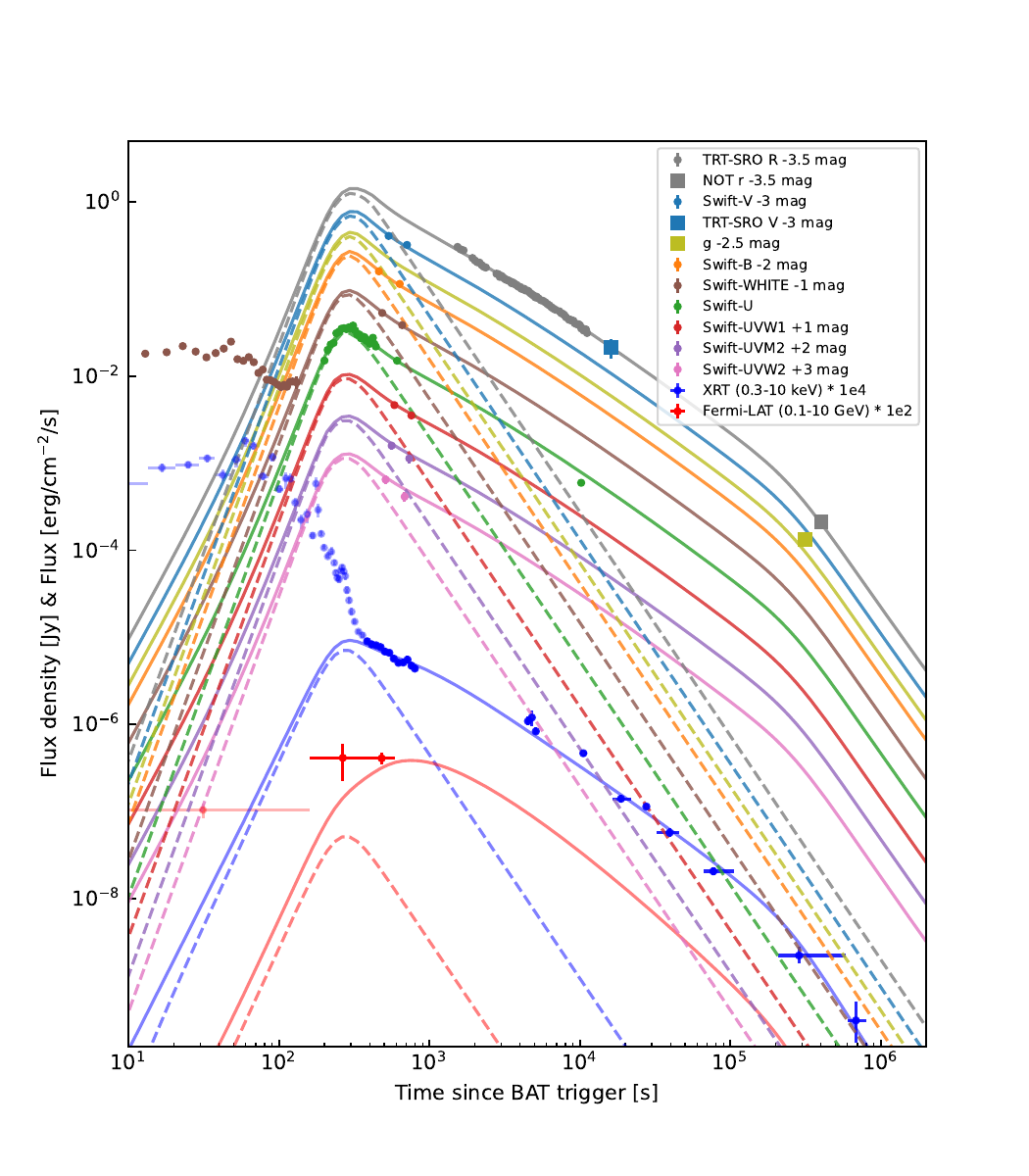}
    \caption{GRB 241030A light curves from XRT, UVOT and Fermi-LAT data. In this scenario, we considers the optical peak as dominated by RS only. Dashed line represents the RS contribution while solid line represents the sum contribution of all components.}
    \label{fig:lc_fs}
\end{figure}

We utilize {\tt PyMultiNest} \citep{2014A&A...564A.125B} to conduct Bayesian inference on the parameters of the afterglow model (Eq. \ref{eq:fs_rs}), setting the number of live points to 500. The prior ranges for these parameters are provided in Table \ref{tab_fsrs_par}. Additionally, we define a log-likelihood term for the each data point, which corresponds to an observation at time $t_i$ in a band with central frequency $\nu_i$, as follows:
\begin{equation}
    \ln \mathcal{L}_i=-\frac{1}{2} \frac{(F_{t, \nu} - F_{t, \nu,\rm obs})^2}{\sigma_i^2+f_{sys}^2F_{t, \nu,\rm obs}^2} - \ln [2 \pi (\sigma_i^2+f_{sys}^2F_{t, \nu,\rm obs}^2)].
\end{equation}
where $f_{sys}$ is a free parameter to characterize the systematic error. By fitting the light curves, we are able to constrain various parameters for both the FS and RS components. The best simultaneous fit results are shown in Figure \ref{fig:lc_fs} and parameters are summarized in Figure \ref{fig:6}. According to the FS and RS models, we estimate that the initial Lorentz factor $\Gamma_0 \sim 135$, the half-opening angle $\theta_j \sim5^{\circ}$. All the parameters are constrained well, except the the isotropic kinetic energy $E_{\rm k,iso}\sim 10^{56}$ erg. The system error here is set as free to account for the bias of the central wavelength of the UVOT WHITE band, since the width of the WHITE band is so large that the reference value could be inaccurate. The fit also constrain well the fraction of the the forward shock energy converted into the energy of magnetic field $\epsilon_{B,f}=2\times10^{-6}$ and the reverse shock energy converted into the energy of magnetic field $\epsilon_{B,r}=7\times10^{-4}$, i.e., the reverse shock region is moderately magnetized, as found in GRB 990123 and other events \citep{2002ChJAA...2..449F,2003ApJ...595..950Z}. 

\section{SUMMARY AND DISCUSSION} \label{sec:summary}
We presented a detailed analysis of the prompt emission and the early afterglow of GRB 241030A, and suggest GRB 241030A is a burst triggered by its precursor emission about 100\,s before the prompt emission. The prompt UV/optical light curve has several bumps, which roughly traces the prompt $\gamma$-ray emission. Shortly after the prompt emission, the optical light curve rises steeply and reaches a bright peak of U = 13.6 AB mag around 410 s, which is attributed to the RS emission of the afterglow. At the same time, the X-ray keeps decaying except for a flare around 278\,s. After 410\,s, both the UV/optical and X-ray light curves show normal decay behaviors. In this work, our main results are as follows:

(i) Because GRB 241030A was triggered by its precursor emission, {\it Swift}-XRT/UVOT were able to observe the prompt emission at a high temporal resolution. According to simultaneous multi-band observations, we find that emissions from UV/optical to $\gamma$-ray show similar complicated overlapped pulses, which suggests their may share the same origin. After the prompt phase, a U-band bump at $\sim410$\,s with a steep rising was observed, which is attributed to the onset of the afterglow. The comprehensive data of GRB 241030A in the prompt phase and the early afterglow phase makes it possible to study the mechanisms of emission and to constrain several parameters of the standard afterglow model.

(ii) In the prompt emission phase, the {\it Swift}/XRT, {\it Swift}/BAT and Fermi-GBM data were jointly fitted. We found that a thermal component is required to account for the excess in the X-ray band. Following \citet{Pe’er_2007}, the upper limit of Lorentz factor of the photosphere , $\Gamma_{\rm ph}$, range between approximately 20 and 80 during the prompt phase. Considering in Pe'er's method, the plasma in the classical fireball has already reached the saturation radius, resulting in the Lorentz factor of the outflow attaining its maximum value. If the outflow were still in the acceleration phase, $\Gamma_{\rm ph}$ might be inaccurate derived from this method. 

(iii) Considering the steep rise($\sim t^{5.3}$) observed in the U band, we adopt the scenario that both the RS and the FS considerably contribute to the afterglow at the early stage. Due to the well-sampled data of the afterglow at the early phase, some physical parameters of RS and FS can be well constrained. The best fitted model shows that the UV/optical light curves are initially dominated by the RS and the FS takes over after $\sim410$\,s. From RS and FS modeling, we have the initial Lorentz factor $\Gamma_0$ $\sim$ 135, the half-opening angle $\theta_j \sim5^{\circ}$ and isotropic kinetic energy $E_{\rm k,iso}\sim 10^{56}$ erg, and $\mathcal{R}_{B}=(\epsilon_{B,r}/\epsilon_{B,f})^{\frac{1}2{}}\approx 20$. Therefore, the GRB ejecta are collimated, with a small opening angle $ \sim5^{\circ}$. Furthermore, the RS region is moderately magnetized, similar to what is found in the literature \citep{2002ChJAA...2..449F,2003ApJ...595..950Z}, which may point towards a magnetized central engine.

In summary,  with broadband and high temporal resolution observations of GRB 241030A, we find emission from the photosphere, the internal shock and the forward and reverse external shock, and their parameters are constrained.


\vspace{5mm}
% \facilities{HST(STIS), Swift(XRT and UVOT), AAVSO, CTIO:1.3m,


% \software{astropy \citepp{2013A&A...558A..33A,2018AJ....156..123A},  
%           Cloudy \citepp{2013RMxAA..49..137F}, 
%           Source Extractor \citepp{1996A&AS..117..393B}
%           }

\begin{acknowledgments}
We thank Prof. Yi-Zhong Fan for the stimulating discussion. This work is supported by the Natural Science Foundation of China (NSFC) under grant Nos. 12321003, 12225305, 12073080, the Strategic Priority Research Program of the Chinese Academy of Sciences, grant No. XDB0550400. Y.W is supported by the Jiangsu Funding Program for Excellent Postdoctoral Talent (grant No. 2024ZB110), the Postdoctoral Fellowship Program (grant No. GZC20241916) and the General Fund of the China Postdoctoral Science Foundation(grant No. 2024M763531). J. R is support by the General Fund (grant No. 2024M763530). We acknowledge the use of the public data from Swift archive, Mikulski Archive for Space Telescopes and CALSPEC. This research is based on observations under program ID TRTToO\_2024005 made with the Thai Robotic Telescope, which is operated by the National Astronomical Research Institute of Thailand (Public Organization).
\end{acknowledgments}

\facilities{Swift(BAT, XRT and UVOT), Fermi(LAT and GBM)}



\newpage
\appendix

In this appendix, we provide detailed information on the observation data and fitting results of the parameters in prompt and afterglow phases.

\begin{ThreePartTable}
\begin{TableNotes}\footnotesize
    \item[]{\bf Note.} The photometry results presented in this table have not been corrected for Galactic extinction.
\end{TableNotes}
\begin{longtable}[htbp]{c c c c c }
\caption{{\it Swift} UVOT photomertric observation\label{table:1}}\\

\hline
$T-T_{0}(\rm s)$ & Exposure(s) & $F_{\nu}$(Jy) & Error(Jy) & Filter\\ 
\hline
\endfirsthead

\hline
$T-T_{0}(s)$ & Exposure(s) & $F_{\nu}$(Jy) & Error(Jy) & Filter\\ 
\hline
\endhead

\hline
\endfoot

\hline
\multicolumn{5}{c}{End of the Table}\\
\hline
\insertTableNotes\\
\endlastfoot

68 &	10 &	$1.293 \times 10^{-3}$ &	$2.270 \times 10^{-4}$ & V \\
88 &	5 &	$1.200 \times 10^{-3}$ &	$8.265 \times 10^{-5}$ &	WHITE \\
93 &	5 &	$1.185 \times 10^{-3}$ &	$8.214 \times 10^{-5}$ &	WHITE \\
98 &	5 &	$1.136 \times 10^{-3}$ &	$7.939 \times 10^{-5}$ &	WHITE \\
103 &	5 &	$1.045 \times 10^{-3}$ &	$7.497 \times 10^{-5}$ &	WHITE \\
108 &	5 &	$1.146 \times 10^{-3}$ &	$8.006 \times 10^{-5}$ &	WHITE \\
113 &	5 &	$1.664 \times 10^{-3}$ &	$1.069 \times 10^{-4}$ &	WHITE \\
118 &	5 &	$1.728 \times 10^{-3}$ &	$1.103 \times 10^{-4}$ &	WHITE \\
123&	5 &	$2.032 \times 10^{-3}$ &	$1.273 \times 10^{-4}$ &	WHITE \\
128 &	5 &	$1.758 \times 10^{-3}$ &	$1.120 \times 10^{-4}$ &	WHITE \\
133 &	5 &	$1.509 \times 10^{-3}$ &	$9.872 \times 10^{-5}$ &	WHITE \\
138 &	5 &	$1.691 \times 10^{-3}$ &	$1.084 \times 10^{-4}$ &	WHITE \\
143 &	5 &	$1.902 \times 10^{-3}$ &	$1.197 \times 10^{-4}$ &	WHITE \\
148 &	5 &	$2.285 \times 10^{-3}$ &	$1.425 \times 10^{-4}$ &	WHITE \\
153 &	5 &	$1.436 \times 10^{-3}$ &	$9.476 \times 10^{-5}$ &	WHITE \\
158 &	5 &	$1.378 \times 10^{-3}$ &	$9.184 \times 10^{-5}$ &	WHITE \\
163 &	5 &	$1.518 \times 10^{-3}$ &	$9.910 \times 10^{-5}$ &	WHITE \\
168 &	5 &	$1.328 \times 10^{-3}$ &	$8.922 \times 10^{-5}$ &	WHITE \\
173 &	5 &	$9.998 \times 10^{-4}$ &	$7.276 \times 10^{-5}$ &	WHITE \\
178 &	5 &	$1.091 \times 10^{-3}$ &	$7.730 \times 10^{-5}$ &	WHITE \\
183 &	5 &	$8.365 \times 10^{-4}$ &	$6.469 \times 10^{-5}$ &	WHITE \\
188 &	5 &	$8.245 \times 10^{-4}$ &	$6.395 \times 10^{-5}$ &	WHITE \\
193 &	5 &	$7.855 \times 10^{-4}$ &	$6.186 \times 10^{-5}$ &	WHITE \\
198 &	5 &	$7.329 \times 10^{-4}$ &	$5.940 \times 10^{-5}$ &	WHITE \\
203 &	5 &	$6.894 \times 10^{-4}$ &	$5.724 \times 10^{-5}$ &	WHITE \\
208 &	5 &	$7.281 \times 10^{-4}$ &	$5.911 \times 10^{-5}$ &	WHITE \\
213 &	5 &	$6.951 \times 10^{-4}$ &	$5.763 \times 10^{-5}$ &	WHITE \\
218 &	5 &	$7.865 \times 10^{-4}$ &	$6.187 \times 10^{-5}$ &	WHITE \\
223 &	5 &	$7.976 \times 10^{-4}$ &	$6.245 \times 10^{-5}$ &	WHITE \\
228 &	5 &	$7.905 \times 10^{-4}$ &	$6.211 \times 10^{-5}$ &	WHITE \\
300 &	10 &	$5.044 \times 10^{-3}$ &	$2.700 \times 10^{-4}$ &	U \\
310 &	10 &	$6.677 \times 10^{-3}$ &	$3.344 \times 10^{-4}$ &	U \\
320 &	10 &	$7.704 \times 10^{-3}$ &	$3.751 \times 10^{-4}$ &	U \\
330 &	10 &	$8.042 \times 10^{-3}$ &	$3.894 \times 10^{-4}$ &	U \\
340 &	10 &	$1.040 \times 10^{-2}$ &	$4.886 \times 10^{-4}$ &	U \\
350 &	10 &	$1.083 \times 10^{-2}$ &	$5.071 \times 10^{-4}$ &	U \\
360 &	10 &	$1.172 \times 10^{-2}$ &	$5.473 \times 10^{-4}$ &	U \\
370 &	10 &	$1.186 \times 10^{-2}$ &	$5.541 \times 10^{-4}$ &	U \\
380 &	10 &	$1.158 \times 10^{-2}$ &	$5.410 \times 10^{-4}$ &	U \\
390 &	10 &	$1.212 \times 10^{-2}$ &	$5.660 \times 10^{-4}$ &	U \\
400 &	10 &	$1.171 \times 10^{-2}$ &	$5.470 \times 10^{-4}$ &	U \\
410 &	10 &	$1.271 \times 10^{-2}$ &	$5.936 \times 10^{-4}$ &	U \\
420 &	10 &	$1.119 \times 10^{-2}$ &	$5.236 \times 10^{-4}$ &	U \\
430 &	10 &	$1.008 \times 10^{-2}$ &	$4.747 \times 10^{-4}$ &	U \\
440 &	10 &	$1.008 \times 10^{-2}$ &	$4.748 \times 10^{-4}$ &	U \\
450 &	10 &	$9.084 \times 10^{-3}$ &	$4.319 \times 10^{-4}$ &	U \\
460 &	10 &	$9.505 \times 10^{-3}$ &	$4.500 \times 10^{-4}$ &	U \\
470 &	10 &	$9.350 \times 10^{-3}$ &	$4.436 \times 10^{-4}$ &	U \\
480 &	10 &	$8.519 \times 10^{-3}$ &	$4.086 \times 10^{-4}$ &	U \\
490 &	10 &	$8.615 \times 10^{-3}$ &	$4.127 \times 10^{-4}$ &	U \\
500 &	10 &	$7.875 \times 10^{-3}$ &	$3.823 \times 10^{-4}$ &	U \\
510 &	10 &	$8.314 \times 10^{-3}$ &	$4.003 \times 10^{-4}$ &	U \\
520 &	10 &	$9.211 \times 10^{-3}$ &	$4.376 \times 10^{-4}$ &	U \\
530 &	10 &	$7.840 \times 10^{-3}$ &	$3.810 \times 10^{-4}$ &	U \\
540 &	9 &	$7.299 \times 10^{-3}$ &	$3.749 \times 10^{-4}$ &	U \\
561 &	19 &	$1.053 \times 10^{-2}$ &	$3.889 \times 10^{-4}$ &	B \\
585 &	19 &	$4.882 \times 10^{-3}$ &	$2.048 \times 10^{-4}$ &	WHITE  \\
611 &	19 &	$5.769 \times 10^{-4}$ &	$6.199 \times 10^{-5}$ &	UVW2  \\
635 &	19 &	$1.176 \times 10^{-2}$ &	$5.618 \times 10^{-4}$ &	V  \\
660 &	18 &	$1.159 \times 10^{-3}$ &	$1.115 \times 10^{-4}$ &	UVM2  \\
684 &	18 &	$2.092 \times 10^{-3}$ &	$1.333 \times 10^{-4}$ &	UVW1  \\
709 &	19 &	$5.024 \times 10^{-3}$ &	$2.089 \times 10^{-4}$ &	U  \\
734 &	19 &	$7.542 \times 10^{-3}$ &	$2.980 \times 10^{-4}$ &	B  \\
758 &	19 &	$3.540 \times 10^{-3}$ &	$1.308 \times 10^{-4}$ &	WHITE  \\
784 &	19 &	$3.665 \times 10^{-4}$ &	$4.896 \times 10^{-5}$ &	UVW2  \\
804 &	19 &	$9.264 \times 10^{-3}$ &	$4.788 \times 10^{-4}$ &	V \\
834 &	17 &	$8.260 \times 10^{-4}$ &	$9.506 \times 10^{-5}$ &	UVM2  \\
858 &	18 &	$1.593 \times 10^{-3}$ &	$1.140 \times 10^{-4}$ &	UVW1  \\
10294 & 300 & $1.991 \times 10^{-4}$ &	$9.556 \times 10^{-6}$ &	U  \\

\end{longtable}
\end{ThreePartTable}

\newpage
\begin{ThreePartTable}
\begin{TableNotes}\footnotesize
    \item[]{\bf Note.} The photometry results presented in this table have not been corrected for Galactic extinction.
\end{TableNotes}

\begin{longtable}[htbp]{c c c c c c}
\caption{Ground-base photomertric observation\label{table:ground obs}}\\
%% \tablecomments{The photometry results presented in this table have not been corrected for Galactic extinction.}
\hline
$T_{mid}-T_{0}(\rm s)$ & Exposure(s) & Filter & Mag (AB) & Error(AB) & Ins.\\
\hline
\endfirsthead
\hline
$T_{mid}-T_{0}(s)$ & Exposure(s) & Filter & Mag (AB) & Error & Ins.\\ 
\hline
\endhead
\hline
\endfoot
\hline
\multicolumn{6}{c}{End of the Table}\\
\hline
\insertTableNotes\\
\endlastfoot

1636 & 60 & R & 14.77 & 0.01 & TRT-SRO\\
1710 & 60 & R & 14.84 & 0.01 & TRT-SRO\\
1783 & 60 & R & 14.87 & 0.01 & TRT-SRO\\
2031 & 60 & R & 15.09 & 0.01 & TRT-SRO\\
2105 & 60 & R & 15.16 & 0.01 & TRT-SRO\\
2177 & 60 & R & 15.21 & 0.01 & TRT-SRO\\
2250 & 60 & R & 15.22 & 0.01 & TRT-SRO\\
2322 & 60 & R & 15.29 & 0.01 & TRT-SRO\\
2395 & 60 & R & 15.34 & 0.01 & TRT-SRO\\
2468 & 60 & R & 15.36 & 0.01 & TRT-SRO\\
2889 & 60 & R & 15.54 & 0.01 & TRT-SRO\\
2962 & 60 & R & 15.55 & 0.01 & TRT-SRO\\
3034 & 60 & R & 15.63 & 0.01 & TRT-SRO\\
3106 & 60 & R & 15.6 & 0.01 & TRT-SRO\\
3177 & 60 & R & 15.64 & 0.01 & TRT-SRO\\
3250 & 60 & R & 15.68 & 0.01 & TRT-SRO\\
3321 & 60 & R & 15.71 & 0.01 & TRT-SRO\\
3394 & 60 & R & 15.72 & 0.01 & TRT-SRO\\
3467 & 60 & R & 15.74 & 0.01 & TRT-SRO\\
3538 & 60 & R & 15.76 & 0.01 & TRT-SRO\\
3610 & 60 & R & 15.81 & 0.01 & TRT-SRO\\
3682 & 60 & R & 15.79 & 0.01 & TRT-SRO\\
3753 & 60 & R & 15.82 & 0.01 & TRT-SRO\\
3825 & 60 & R & 15.85 & 0.01 & TRT-SRO\\
3897 & 60 & R & 15.86 & 0.01 & TRT-SRO\\
3968 & 60 & R & 15.9 & 0.01 & TRT-SRO\\
4040 & 60 & R & 15.92 & 0.01 & TRT-SRO\\
41125 & 60 & R & 15.92 & 0.01 & TRT-SRO\\
4185 & 60 & R & 15.97 & 0.01 & TRT-SRO\\
4257 & 60 & R & 15.98 & 0.01 & TRT-SRO\\
4344 & 90 & R & 15.97 & 0.01 & TRT-SRO\\
4447 & 90 & R & 16.02 & 0.01 & TRT-SRO\\
4550 & 90 & R & 16.02 & 0.01 & TRT-SRO\\
4652 & 90 & R & 16.04 & 0.01 & TRT-SRO\\
4755 & 90 & R & 16.09 & 0.01 & TRT-SRO\\
4857 & 90 & R & 16.12 & 0.01 & TRT-SRO\\
4958 & 90 & R & 16.15 & 0.01 & TRT-SRO\\
5061 & 90 & R & 16.18 & 0.01 & TRT-SRO\\
5162 & 90 & R & 16.22 & 0.01 & TRT-SRO\\
5265 & 90 & R & 16.2 & 0.01 & TRT-SRO\\
5413 & 180 & R & 16.25 & 0.01 & TRT-SRO\\
5606 & 180 & R & 16.3 & 0.01 & TRT-SRO\\
5798 & 180 & R & 16.32 & 0.01 & TRT-SRO\\
5990 & 180 & R & 16.36 & 0.01 & TRT-SRO\\
6182 & 180 & R & 16.41 & 0.01 & TRT-SRO\\
6375 & 180 & R & 16.45 & 0.01 & TRT-SRO\\
6567 & 180 & R & 16.5 & 0.01 & TRT-SRO\\
6758 & 180 & R & 16.54 & 0.01 & TRT-SRO\\
6951 & 180 & R & 16.57 & 0.01 & TRT-SRO\\
7410 & 180 & R & 16.62 & 0.01 & TRT-SRO\\
7601 & 180 & R & 16.67 & 0.01 & TRT-SRO\\
7793 & 180 & R & 16.69 & 0.01 & TRT-SRO\\
7986 & 180 & R & 16.73 & 0.01 & TRT-SRO\\
8179 & 180 & R & 16.79 & 0.01 & TRT-SRO\\
8565 & 180 & R & 16.86 & 0.01 & TRT-SRO\\
8758 & 180 & R & 16.86 & 0.01 & TRT-SRO\\
8950 & 180 & R & 16.89 & 0.01 & TRT-SRO\\
9143 & 180 & R & 16.89 & 0.02 & TRT-SRO\\
9336 & 180 & R & 16.96 & 0.02 & TRT-SRO\\
9528 & 180 & R & 16.98 & 0.02 & TRT-SRO\\
9721 & 180 & R & 16.99 & 0.02 & TRT-SRO\\
9914 & 180 & R & 16.99 & 0.02 & TRT-SRO\\
10106 & 180 & R & 17.02 & 0.03 & TRT-SRO\\
10299 & 180 & R & 17.11 & 0.03 & TRT-SRO\\
10491 & 180 & R & 17.11 & 0.04 & TRT-SRO\\
10682 & 180 & R & 17.17 & 0.04 & TRT-SRO\\
10874 & 180 & R & 17.13 & 0.04 & TRT-SRO\\
11067 & 180 & R & 17.14 & 0.04 & TRT-SRO\\
11254 & 180 & R & 17.24 & 0.04 & TRT-SRO\\
15767 & 3x150 & B & $>$ 18.1 &  & TRT-SRO\\
16254 & 3x150 & V & 18.14 & 0.27 & TRT-SRO\\
16953 & 3x150 & R & $>$ 17.6 &  & TRT-SRO\\
17442 & 3x150 & I & $>$ 15.7 &  & TRT-SRO\\
312924 & 3x300 & g & 23.08 & 0.13 & NOT\\
314248 & 5x300 & z & $>$ 21.8 &  & NOT\\
400777 & 3x600 & r & 23.17 & 0.12 & NOT\\
\end{longtable}
\end{ThreePartTable}

\newpage
\begin{table}[t]
\centering
\begin{tabular}{c c c c c c c c c c} 
 \hline
 Data & Model & start [s] & stop [s] & $\alpha$ & $E_{peak}$ [keV]& norm & kT [keV]& norm$_{BB}$ & $\Delta_{\rm BIC}$\\ 
 \hline
XRT+BAT+GBM &	PL+BB &	85 &	90 &	$-1.45^{+0.03}_{-0.03}$ &	 &	$1.33_{-0.15}^{+0.15}$&	$2.58_{-0.18}^{+0.18}$ &	$0.26_{-0.04}^{+0.04}$ &	31.06\\
XRT+BAT &	PL+BB &	90 &	95 &	$-1.77^{+0.07}_{-0.07}$ &	 &	$1.78_{-0.37}^{+0.35}$&	$1.40_{-0.24}^{+0.39}$ &	$0.18_{-0.04}^{+0.07}$ &	36.48\\
XRT+BAT &	PL+BB &	95 &	100 &	$-1.70^{+0.16}_{-0.13}$ &	 &	$0.75_{-0.32}^{+0.33}$&	$0.79_{-0.06}^{+0.08}$ &	$0.10_{-0.02}^{+0.02}$ &	36.47\\
XRT+BAT &	PL+BB &	100 &	105 &	$-1.71^{+0.14}_{-0.11}$ &	 &	$0.87_{-0.34}^{+0.34}$&	$0.82_{-0.07}^{+0.09}$ &	$0.11_{-0.02}^{+0.02}$ &	38.11\\
XRT+BAT+GBM &	CPL+BB &	105 &	110 &	$-1.24^{+0.06}_{-0.05}$ &	$74.63_{-14.10}^{+20.48}$ &	$2.22_{-0.29}^{+0.27}$&	$1.87_{-0.35}^{+0.30}$ &	$0.25_{-0.06}^{+0.06}$ &	22.56\\
XRT+BAT+GBM &	CPL+BB &	110 &	115 &	$-1.02^{+0.03}_{-0.03}$ &	$248.38_{-30.46}^{+36.77}$ &	$2.38_{-0.20}^{+0.20}$&	$3.06_{-0.22}^{+0.27}$ &	$0.59_{-0.07}^{+0.07}$ &	62.80\\
XRT+BAT+GBM &	CPL+BB &	115 &	120 &	$-1.18^{+0.03}_{-0.03}$ &	$129.83_{-10.63}^{+12.18}$ &	$2.49_{-0.21}^{+0.21}$&	$2.17_{-0.16}^{+0.16}$ &	$0.47_{-0.06}^{+0.06}$ &	66.47\\
XRT+BAT+GBM &	CPL+BB &	120 &	125 &	$-1.08^{+0.03}_{-0.03}$ &	$193.98_{-22.43}^{+26.72}$ &	$3.01_{-0.26}^{+0.26}$&	$2.63_{-0.15}^{+0.16}$ &	$0.84_{-0.08}^{+0.08}$ &	105.36\\
XRT+BAT+GBM &	CPL+BB &	125 &	130 &	$-1.24^{+0.05}_{-0.04}$ &	$63.53_{-8.74}^{+11.21}$ &	$3.39_{-0.37}^{+0.37}$&	$1.67_{-0.18}^{+0.18}$ &	$0.45_{-0.07}^{+0.07}$ &	56.06\\
XRT+BAT+GBM &	CPL+BB &	130 &	135 &	$-1.16^{+0.02}_{-0.02}$ &	$158.16_{-16.13}^{+19.18}$ &	$5.17_{-0.36}^{+0.36}$&	$3.28_{-0.27}^{+0.32}$ &	$0.58_{-0.08}^{+0.09}$ &	42.99\\
XRT+BAT+GBM &	CPL+BB &	135 &	140 &	$-1.15^{+0.02}_{-0.02}$ &	$402.90_{-60.35}^{+78.04}$ &	$4.22_{-0.30}^{+0.29}$&	$3.01_{-0.59}^{+0.80}$ &	$0.35_{-0.09}^{+0.08}$ &	11.80\\
XRT+BAT+GBM &	CPL+BB &	140 &	145 &	$-1.17^{+0.02}_{-0.02}$ &	$431.21_{-66.43}^{+86.15}$ &	$3.90_{-0.30}^{+0.30}$&	$2.33_{-0.34}^{+0.35}$ &	$0.25_{-0.07}^{+0.07}$ &	2.97\\
XRT+BAT+GBM &	CPL+BB &	145 &	150 &	$-0.82^{+0.04}_{-0.04}$ &	$150.49_{-10.86}^{+11.80}$ &	$2.19_{-0.25}^{+0.26}$&	$1.45_{-0.14}^{+0.15}$ &	$0.29_{-0.06}^{+0.06}$ &	17.91\\
XRT+BAT+GBM &	CPL+BB &	150 &	155 &	$-1.19^{+0.04}_{-0.04}$ &	$63.06_{-6.94}^{+8.57}$ &	$3.90_{-0.34}^{+0.34}$&	$2.77_{-0.15}^{+0.16}$ &	$0.73_{-0.08}^{+0.08}$ &	76.81\\
XRT+BAT+GBM &	CPL+BB &	155 &	160 &	$-1.34^{+0.04}_{-0.04}$ &	$80.01_{-12.99}^{+18.38}$ &	$5.70_{-0.49}^{+0.49}$&	$2.41_{-0.15}^{+0.16}$ &	$0.99_{-0.09}^{+0.09}$ &	112.17\\
XRT+BAT+GBM &	CPL+BB &	160 &	165 &	$-1.05^{+0.03}_{-0.03}$ &	$139.80_{-9.50}^{+10.52}$ &	$5.96_{-0.47}^{+0.48}$&	$2.13_{-0.19}^{+0.23}$ &	$0.68_{-0.11}^{+0.11}$ &	26.42\\
XRT+BAT+GBM &	CPL+BB &	165 &	170 &	$-1.10^{+0.04}_{-0.03}$ &	$78.20_{-5.44}^{+6.10}$ &	$5.69_{-0.54}^{+0.55}$&	$2.00_{-0.24}^{+0.28}$ &	$0.56_{-0.11}^{+0.11}$ &	17.93\\
XRT+BAT+GBM &	CPL+BB &	170 &	175 &	$-1.25^{+0.10}_{-0.08}$ &	$36.47_{-5.34}^{+6.56}$ &	$3.11_{-0.66}^{+0.64}$&	$1.08_{-0.09}^{+0.11}$ &	$0.46_{-0.05}^{+0.06}$ &	71.43\\
XRT+BAT &	PL+BB &	175 &	180 &	$-1.98^{+0.07}_{-0.07}$ &	 &	$3.66_{-0.73}^{+0.70}$&	$1.04_{-0.11}^{+0.15}$ &	$0.33_{-0.04}^{+0.04}$ &	73.46\\
XRT+BAT &	CPL+BB &	180 &	185 &	$-1.63^{+0.09}_{-0.08}$ &	$23.30_{-7.26}^{+14.21}$ &	$4.35_{-0.66}^{+0.61}$&	$1.10_{-0.18}^{+0.28}$ &	$0.19_{-0.04}^{+0.04}$ &	13.93\\
XRT+BAT+GBM &	CPL+BB &	185 &	190 &	$-1.33^{+0.03}_{-0.03}$ &	$96.20_{-12.80}^{+16.21}$ &	$6.31_{-0.52}^{+0.52}$&	$2.15_{-0.20}^{+0.22}$ &	$0.56_{-0.09}^{+0.09}$ &	30.63\\
XRT+BAT+GBM &	CPL+BB &	190 &	195 &	$-1.33^{+0.10}_{-0.09}$ &	$17.00_{-3.09}^{+4.07}$ &	$4.02_{-0.61}^{+0.57}$&	$1.24_{-0.13}^{+0.14}$ &	$0.46_{-0.06}^{+0.07}$ &	47.52\\
XRT+BAT &	PL+BB &	195 &	200 &	$-1.82^{+0.10}_{-0.09}$ &	 &	$2.08_{-0.64}^{+0.65}$&	$0.79_{-0.05}^{+0.06}$ &	$0.34_{-0.03}^{+0.03}$ &	112.81\\
XRT+BAT &	PL+BB &	200 &	205 &	$-2.26^{+0.15}_{-0.21}$ &	 &	$2.33_{-0.64}^{+0.61}$&	$0.73_{-0.07}^{+0.09}$ &	$0.19_{-0.02}^{+0.03}$ &	49.68\\
XRT+BAT &	PL+BB &	205 &	210 &	$-1.78^{+0.05}_{-0.05}$ &	 &	$3.75_{-0.62}^{+0.60}$&	$0.98_{-0.13}^{+0.19}$ &	$0.18_{-0.03}^{+0.03}$ &	25.36\\
XRT+BAT &	PL+BB &	210 &	215 &	$-1.94^{+0.05}_{-0.05}$ &	 &	$4.77_{-0.66}^{+0.65}$&	$1.04_{-0.09}^{+0.11}$ &	$0.36_{-0.04}^{+0.04}$ &	82.08\\
XRT+BAT &	CPL+BB &	215 &	220 &	$-1.60^{+0.06}_{-0.06}$ &	$47.27_{-13.17}^{+26.15}$ &	$5.05_{-0.65}^{+0.62}$&	$1.26_{-0.17}^{+0.23}$ &	$0.30_{-0.05}^{+0.06}$ &	29.94\\
XRT+BAT &	PL+BB &	220 &	225 &	$-1.86^{+0.04}_{-0.04}$ &	 &	$5.11_{-0.63}^{+0.61}$&	$1.13_{-0.15}^{+0.22}$ &	$0.28_{-0.04}^{+0.05}$ &	52.32\\
XRT+BAT &	PL+BB &	225 &	230 &	$-1.98^{+0.06}_{-0.06}$ &	 &	$3.05_{-0.47}^{+0.45}$&	$0.84_{-0.09}^{+0.12}$ &	$0.14_{-0.02}^{+0.02}$ &	38.05\\

\hline
\end{tabular}
\caption{Best spectral fit result from XRT, BAT, and GBM. Joint fit obtained using either with the PL+BB or CPL+BB model. Spectral normalizations are computed in $\rm ph\cdot cm^{-2}\cdot s^{-1}$ at 1keV. Here, $\alpha$ is the photon index of PL or CPL, $\Delta$ BIC is differences of BIC value between model with BB and model without BB. Except for the epoch at 140 - 145 seconds, where $\Delta_{\rm BIC}$ = 2.97, we find $\Delta_{\rm BIC} > 10$ in all other epochs and $\Delta_{\rm BIC}>30 $ in 20 of 29 epochs, which is a strong evidence of an additional thermal component.}
\label{table:prompt_fit}
\end{table}

\begin{table}[htbp]
\centering
\caption{SED Data at $T_{0}+755 s$}
\begin{tabular}{c c c } 
 \hline
 Filter & $F_{\nu}$(Jy) & Error(Jy)\\ 
 \hline
 WHITE & $3.540\times 10^{-3}$ & $1.308 \times 10^{-4}$  \\ 
 V & $1.020\times 10^{-2}$ & $5.274\times 10^{-4}$  \\
 U & $4.602\times 10^{-3}$ & $1.913\times 10^{-4}$  \\
 B & $7.250\times 10^{-3}$ & $2.865\times 10^{-4}$  \\
 UVW1 & $1.822\times 10^{-3}$ & $1.162\times 10^{-4}$  \\
 UVW2 & $4.291\times 10^{-4}$ & $4.611\times 10^{-5}$  \\
 UVM2 & $9.604\times 10^{-4}$ & $9.240\times 10^{-5}$  \\[1ex] 
 \hline
\end{tabular}
\label{table:2}
\end{table}

\begin{deluxetable}{ccc}[htbp]
\label{tab_fsrs_par}
%\tabletypesize{\tiny}
\tablecaption{Fitting results of model parameters}
\tablehead{
\colhead{Parameter} & \colhead{Prior range} & \colhead{Posterior value}}
\startdata
$\log_{10}\Gamma_{0}$    & $[1,2.5]$  & $2.13^{+0.04}_{-0.04}$ \\
$\log_{10}\epsilon_{e,f}$  & $[-3,-0.1]$   & $-1.87^{+0.06}_{-0.05}$ \\
$\log_{10}\epsilon_{B,f}$  & $[-7,-0.1]$   & $-5.74^{+0.07}_{-0.07}$ \\
$\log_{10}\theta_{j}$  (rad)    & $[-2,0]$   & $-1.10^{+0.04}_{-0.04}$ \\
$\log_{10}E_{\rm k,iso}$ (erg) & $[51,56]$  & $55.83^{+0.09}_{-0.09}$ \\
$p_{f}$             & $[2,3]$   &  $2.77^{+0.01}_{-0.01}$ \\
\hline
$\log_{10}\epsilon_{e,r}$  & $[-3,-0.1]$ & $-3.15^{+0.09}_{-0.09}$ \\
$\log_{10}\epsilon_{B,r}$  & $[-7,-0.1]$ & $-3.19^{+0.02}_{-0.01}$ \\
$p_{r}$           & $[2,3]$ & $2.09^{+0.05}_{-0.03}$ \\
\hline
$\log_{10}n_{0}$      & $[-3,3]$ & $1.35^{+0.26}_{-0.25}$ \\
%$E(B-V)$$_{\rm host}$  & $[0.3,0.6]$ & $0.38^{+0.02}_{-0.02}$\\
$\log_{10}f_{sys}$      & $[-2,0]$ & $-1.58^{+0.05}_{-0.04}$ \\
\enddata
%\tablecomments{}
\tablenotetext{a}{Uniform prior distribution.}
\tablenotetext{b}{The confidence interval is set to $1\sigma$.}
\end{deluxetable}

\begin{figure}[htbp]
    \centering
    \includegraphics[width=1.0\textwidth]{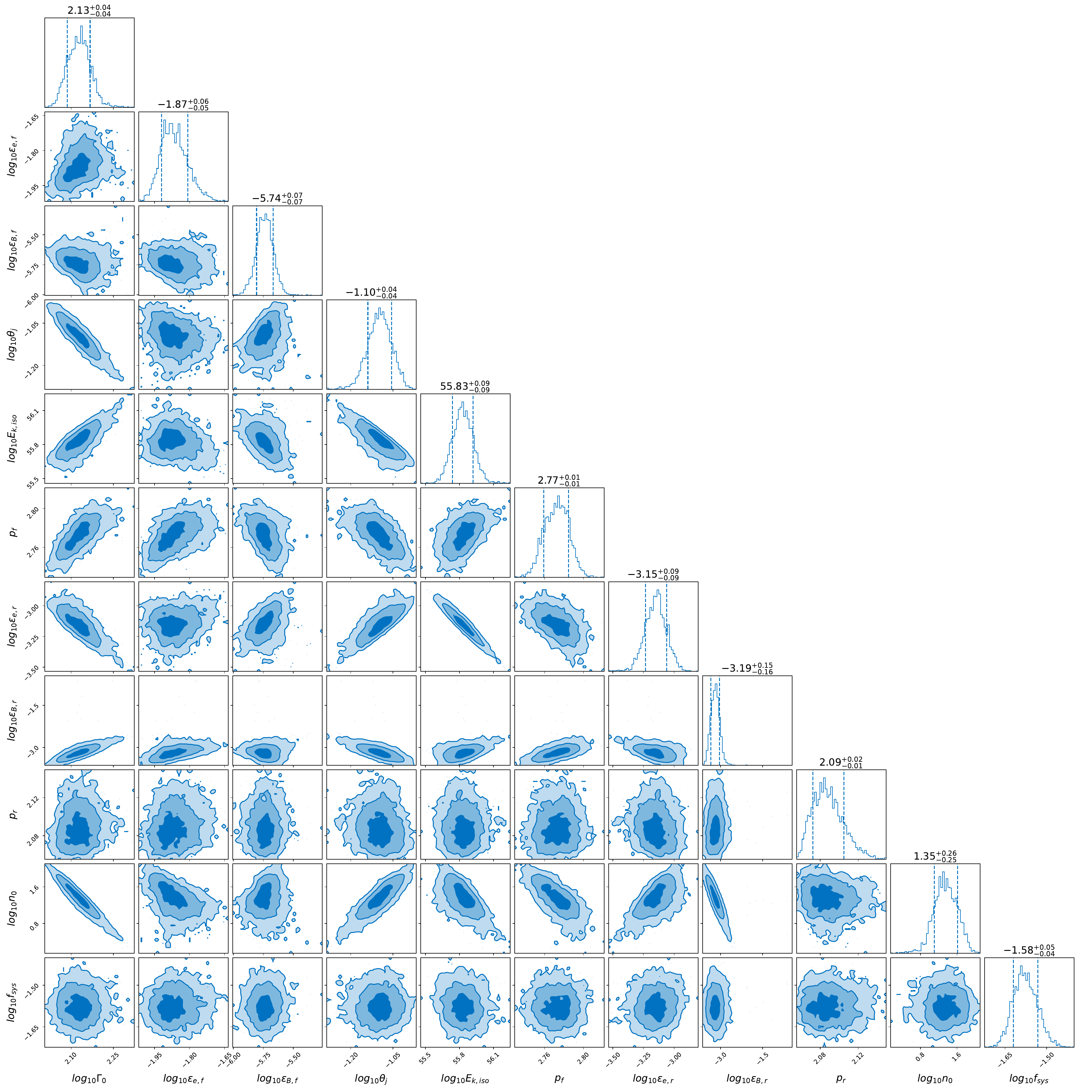}
    \caption{Constrained parameters of the afterglow model for GRB 241030A. Here $\Gamma_0$ is the initial Lorentz factor, $\epsilon_e$ is fraction of energy of relativistic electrons,$\epsilon_b$ is fraction of the energy of magnetic field, $\theta_j$ is the half-opening angle in radians, $E_{\rm k,iso}$ is the isotropic kinetic energy, $p$ is the electron energy distribution index, $n_0$ is the medium number density. And the additional subscripts $f$ and $r$ represent the FS and RS, respectively. }
    \label{fig:6}
\end{figure}

\clearpage
\bibliography{sample631.bib}
\bibliographystyle{aasjournal}



\end{document}